\documentclass{article}
\usepackage{spconf}
\usepackage{cite,graphicx,amsmath,amssymb,booktabs,microtype,array}
\usepackage[table]{xcolor}
\usepackage[hidelinks]{hyperref}
\newcommand{\pg}{\mathrm{PG}}
\newcommand{\pc}{\mathrm{PC}}

\title{Where You Put It Matters: Characterizing
Positional Sensitivity in Large Audio Language Models}
\name{Hui Wang$^{1,2,*}$, Meng Gao$^{2,*}$, Yunfei Chu$^{2,*}$, Jin Xu$^{2,\dagger}$, Yong Qin$^{1,\dagger}$\thanks{$^{*}$Equal contribution.}\thanks{$^{\dagger}$Corresponding authors.}}
\address{$^{1}$College of Computer Science, Nankai University\\
$^{2}$Alibaba Token Hub, Alibaba Group\\
\texttt{wanghui\_hlt@mail.nankai.edu.cn}}
\begin{document}
\maketitle

\begin{abstract}
Multi-audio reasoning requires large audio-language models to compare recordings presented sequentially, making input position a potential confound in evaluation. Ideally, model judgments should depend on acoustic content rather than where a recording appears, yet serialization inevitably changes its position and surrounding context. We study this positional sensitivity with balanced permutations that vary recording order while keeping the comparison problem and correct answer identifier fixed. Our analysis examines three aspects: whether position effects occur systematically, how they vary across tasks and target criteria, and how they change with input scale. Across five models and ten tasks, we observe significant position effects with gaps of 5.8--13.8 percentage points. The effects vary substantially across tasks and targets, and can partially cancel when aggregated. As candidate sets grow from 5 to 32 recordings, accuracy consistently declines, while changes in relative positional effects remain model- and task-dependent. These findings suggest that multi-audio benchmarks should account for positional sensitivity rather than relying on performance under a single input arrangement.
\end{abstract}

\section{Introduction}
Large audio-language models (LALMs) have expanded audio understanding from recognition and transcription to more general reasoning over speech, environmental sounds, and music~\cite{qwen2audio,audioflamingo3,mmau,diffa,uat}. Beyond reasoning about a single recording in isolation, many emerging tasks require models to compare or integrate information across multiple recordings. This shift toward multi-audio reasoning has motivated recent benchmarks that explicitly evaluate understanding across multiple audio inputs~\cite{beyond,mugen,polyaudio,speechllmasjudge}.

Multi-audio understanding, however, introduces a new structural challenge. Although comparisons across recordings should depend only on their acoustic content and the given criterion, the recordings must be serialized into a single input. Consequently, each candidate appears at a different position and in a different context, making an otherwise arbitrary ordering potentially affect the model's judgment.

This raises a basic reliability question for multi-audio evaluation:
\emph{does a model make the same comparison when the correct recording appears at different input positions?} For example, selecting the longest recording
should yield the same answer whether that recording is presented first, in the
middle, or last. If performance varies systematically with position, benchmark
scores may partly reflect input arrangement rather than the underlying
comparison ability being measured.

Order effects have been documented in text multiple-choice evaluation and in LALMs when written or spoken answer choices are reordered~\cite{optionorder,selectionbias,hearing}. These studies largely concern the ordering of response alternatives. Multi-audio reasoning introduces a distinct source of order dependence: the recordings themselves must be presented sequentially. Recent benchmarks such as MUGEN and PolyAudio have established multi-audio reasoning as an important evaluation setting~\cite{mugen,polyaudio}. However, how model behavior depends on the position of individual recordings within a multi-audio sequence remains largely uncharacterized.

We therefore study \emph{multi-audio positional sensitivity}: variation in model behavior induced by changing the input position of the target recording while
keeping the underlying comparison problem unchanged. Our study addresses three questions:
\begin{itemize}
\item \textbf{RQ1 (Existence):} Do systematic position effects persist when
answer identity is held fixed?
\item \textbf{RQ2 (Structure):} How do positional profiles vary across tasks
and target criteria?
\item \textbf{RQ3 (Scaling):} How does positional sensitivity change as multi-audio inputs scale to more candidates and longer contexts?
\end{itemize}
To distinguish positional effects from answer-token identity, we bind each recording
to a neutral identifier and move the identifier together with the recording
across balanced permutations, keeping the correct output fixed while varying
the target recording's position. Across ten MUGEN~\cite{mugen} tasks and five LALMs, we find
systematic positional sensitivity. These effects are not well
described by a single global positional preference: their direction and shape
vary across tasks and target criteria, and can partially cancel when aggregated.
Five-scale comparisons further reveal accuracy losses across sampled positions,
with unequal declines changing relative position advantages.

\section{Measuring Position Sensitivity}\label{sec:protocol}
\subsection{Fixed answer identity and balanced permutations}
Question $i$ contains an instruction $q_i$ and a candidate set of recordings
$\mathcal X_i=\{x_{i,1},\ldots,x_{i,K}\}$ with a unique target
$x_i^\star\in\mathcal X_i$. Each evaluation prompt presents a permutation of
$\mathcal X_i$ as an input sequence. We hold $q_i$, $\mathcal X_i$, and
$x_i^\star$ fixed, varying only the position of $x_i^\star$ in that sequence.
Following symbol--content coupling~\cite{selectionbias}, each recording
receives an identifier that moves with it, keeping the correct output
unchanged. Five-candidate experiments use \texttt{ZK}, \texttt{QV},
\texttt{HN}, \texttt{BF}, and \texttt{JM}, balancing correct labels within tasks.
The instruction precedes the sequence; each identifier precedes its recording.

Five randomized Latin-square permutations per question place every recording
at every position once. Label assignment and permutations use separate random
sources. This balances position exposure, while neighboring recordings and
prefix duration still change.

\subsection{Metrics and statistical principles}
For correctness $y_{i,p}$ of question $i$ at target position $p$, define
\begin{equation}
\begin{aligned}
 A(p)&=\frac{1}{N}\sum_i y_{i,p},\qquad
 \bar A=\frac{1}{|\mathcal P|}\sum_{p\in\mathcal P}A(p),\\
 \pg&=\max_{p\in\mathcal P}A(p)-\min_{p\in\mathcal P}A(p).
\end{aligned}
\end{equation}
Here $\pg$ is the position gap; $\bar A$ is position-balanced accuracy when
$\mathcal P$ covers all positions, otherwise an anchor average. First-slot
contrast $B=A(1)-\operatorname{mean}_{p\in\mathcal P\setminus\{1\}}A(p)$
measures relative first-position advantage. Permutation consistency $\pc$
averages within-question pairwise agreement of predicted recording identities;
invalid predictions contribute zero agreement.

Cochran's $Q$~\cite{cochran} tests equality of accuracy across positions in
single-response sweeps. Bootstrap confidence intervals quantify uncertainty by
resampling questions while preserving paired observations; intervals are
pointwise and unadjusted. Position-profile and interaction tests use 19,999
question-cluster wild-bootstrap draws. Holm correction~\cite{holm} controls
multiple testing within each analysis family.

\subsection{Data and Evaluation Setup}\label{sec:implementation}

We adapt ten MUGEN~\cite{mugen} tasks: keyword spotting, dysarthria, emotion intensity, pitch, duration, speaking rate, signal-to-noise ratio (SNR), interruption, instruments, and maximum-attribute selection. We keep the original five-recording candidate sets and targets, changing only identifiers and permutations. Follow-up experiments recombine the same audio pools using annotations or measured durations. We evaluate 50 questions per task (500 total); all questions have complete cross-model results and receive equal weight.

We evaluate Gemini 3.1 Pro, Gemini 3.5 Flash, Gemini 3.8 Flash\footnote{\url{https://ai.google.dev/gemini-api/docs/models}},
Qwen3.5 Omni Plus\footnote{\url{https://help.aliyun.com/en/model-studio/qwen3-5-omni-plus}}, and Doubao Seed 2.0 Lite\footnote{\url{https://ark.volcengine.com/region:cn-beijing/model/detail?Id=doubao-seed-2-0-lite}}.
Follow-ups use Gemini 3.8 and Qwen3.5, both supporting 32 recordings.
Audio retains its original format without resampling, cropping, or concatenation.
Requests use temperature 0 and at most 2,048 output tokens; other settings use
service defaults. Unparseable answers count as incorrect.

\section{RQ1: Does Candidate Position Matter?}\label{sec:benchmark}

\begin{table}[!t]
\centering
\caption{Ten-task survey. Acc is $\bar A$ (\%); PG and its 95\% bootstrap CI are in percentage points. $p_H$ is Holm-corrected across the five models; PC is permutation consistency.}
\label{tab:main}
\small\setlength{\tabcolsep}{2.5pt}
\renewcommand{\arraystretch}{1.1}
\begin{tabular}{@{}lrrrrr@{}}
\toprule
Model & Acc (\%) & PG & 95\% CI & $p_H$ & PC \\
\midrule
Gemini 3.1 Pro & 67.9 & 5.8 & [4.2, 9.8] & 0.003 & 0.779 \\
Gemini 3.5 Flash & 64.0 & 11.2 & [7.4, 15.2] & $<0.001$ & 0.712 \\
Gemini 3.8 Flash & 66.6 & 7.8 & [5.0, 12.0] & 0.003 & 0.712 \\
Qwen3.5 Omni Plus & 70.7 & 6.2 & [4.4, 9.8] & 0.003 & 0.776 \\
Doubao Seed 2.0 Lite & 65.2 & 13.8 & [10.0, 18.0] & $<0.001$ & 0.647 \\
\bottomrule
\end{tabular}

\end{table}

\begin{figure}[!t]
\centering
\includegraphics[width=\columnwidth]{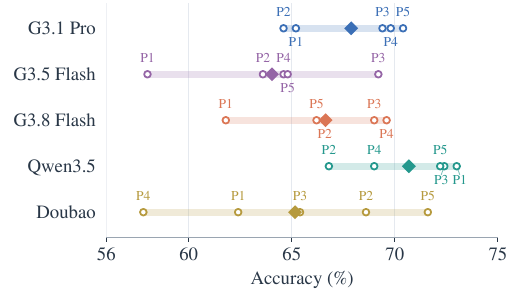}
\caption{Ten-task position accuracies (circles), means (diamonds), and minimum--maximum spans. Labels identify positions P1--P5, combining ties.}
\label{fig:models}
\end{figure}

We first characterize positional sensitivity and assess its robustness to repeat variability and distractor ordering.

\subsection{Main Position Effects}

Table~\ref{tab:main} shows systematic positional sensitivity across all five 
models. Aggregated over the ten tasks, position gaps range from 5.8 to 13.8 
percentage points, with significant effects for every model after Holm correction.

Importantly, the direction of the effect is model-dependent. In 
Fig.~\ref{fig:models}, Qwen3.5 performs best at P1, whereas Gemini 3.8 performs 
worst at the same position. This difference also changes their observed relative 
performance: Qwen3.5's average lead over Gemini 3.8 is about 4 percentage points 
across positions, but widens to 11.2 points at P1. Thus, input position can affect 
both absolute accuracy and observed cross-model comparisons.

\subsection{Controlling for Order}\label{sec:order_control}

As shown in Table~\ref{tab:order_control}, we next test whether the observed position effects can be explained by
run-to-run variability or distractor ordering. For Gemini 3.8 and Qwen3.5
on duration and dysarthria, we evaluate five target positions under five
distractor orders, with two calls per prompt. Identical-prompt repetitions
show 6.8--16.2\% correctness disagreement, while changing distractor order
at a fixed target position introduces a further 1.5--10.8 percentage
points of disagreement; the 95\% CI for Gemini 3.8 on duration includes zero.

Averaging the ten responses for each question--position pair, the position
profiles remain significant across all eight model--task--target conditions,
with Holm-adjusted $p$ values up to 0.0286. Thus, the observed sensitivity
cannot be explained solely by repeated-call variability or any particular
distractor ordering. However, moving the target also changes its neighboring
recordings and the amount of preceding audio, so we interpret the effect as
\emph{position-conditioned sensitivity} rather than a purely absolute-position
effect.

\begin{table}[!t]
\centering
\caption{Repeat and Order correctness disagreement (\%); Excess = Order - Repeat (pp). Rates are averaged within question–position pairs, then across positions.}
\label{tab:order_control}
\small\setlength{\tabcolsep}{4pt}
\renewcommand{\arraystretch}{1.1}
\begin{tabular}{@{}llrrr@{}}
\toprule
Model & Task & Repeat & Order & Excess \\
\midrule
Gemini 3.8 Flash & Duration & 16.2 & 17.8 & +1.5 \\
Gemini 3.8 Flash & Dysarthria & 12.7 & 20.0 & +7.3 \\
Qwen3.5 Omni Plus & Duration & 8.5 & 19.3 & +10.8 \\
Qwen3.5 Omni Plus & Dysarthria & 6.8 & 14.3 & +7.5 \\
\bottomrule
\end{tabular}

\end{table}

\section{RQ2: How Do Effects Vary Across Tasks and Targets?}\label{sec:shape}
Task requirements and target criteria define the context of an audio
comparison. We examine positional sensitivity in relation to these conditions.

\subsection{Task-Level Patterns}

\begin{figure}[!t]
\centering
\includegraphics[width=\columnwidth]{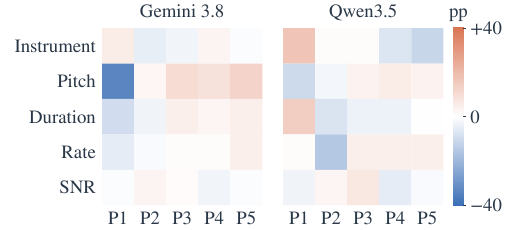}
\caption{Accuracy at each position minus the task mean (pp). Blue is below the mean; orange is above.}
\label{fig:tasks}
\end{figure}

Position effects vary substantially across tasks. Both models reject a
shared position profile across the ten tasks
(Table~\ref{tab:task_interaction}); the result remains significant after
excluding keywords, whose mean accuracy is 98.0--100.0\%
($p_H \leq 0.00010$). Fig.~\ref{fig:tasks} illustrates this heterogeneity.
For example, Gemini 3.8 averages 77.6\% on pitch but drops to 44.0\% at P1, compared with 86.0\% averaged over P2--P5. On speaking rate, Qwen scores 42.0\% at P2 versus 60.0\% averaged over P1 and P5. These pointwise comparisons are
descriptive, but illustrate how task-level averages can conceal
position-specific weaknesses.

Such variation can also affect cross-model comparisons. On pitch,
Qwen's position-balanced lead over Gemini 3.8 is 14.8 percentage points,
but the observed lead is 38.0 points at P1 and 9.0 points across the
remaining positions. The paired model--position test is significant
($p_H=0.0144$, Holm-corrected across ten tasks). Thus, positional
balancing is important not only for characterizing task performance,
but also for making stable comparisons between models.

\begin{table}[!t]
\centering
\caption{Tests of a shared position profile across ten tasks. RMS: question-weighted
deviation from the shared centered profile (pp). $p_H$: Holm correction
across the two models.}
\label{tab:task_interaction}
\small\setlength{\tabcolsep}{3pt}
\renewcommand{\arraystretch}{1.1}
\begin{tabular}{@{}lrr@{}}
\toprule
Model & RMS (pp) & 10 tasks $p_H$ \\
\midrule
Gemini 3.8 Flash & 6.5 & 0.00010 \\
Qwen3.5 Omni Plus & 5.8 & 0.00010 \\
\bottomrule
\end{tabular}

\end{table}

\subsection{Target-Level Patterns}

We further examine whether positional sensitivity varies across target criteria
within the same task. Using the order-controlled duration experiments from
Section~\ref{sec:order_control}, we compare position profiles for questions
targeting the longest versus shortest recording, averaging over five distractor
orders and two calls per prompt.

For Gemini 3.8, the longest and shortest targets have position gaps of 24.4
and 18.1 points, but opposite first-slot contrasts of $-12.3$ and $+15.0$
points. Qwen3.5 shows the same pattern, with position gaps of 15.9 and 21.9
points and first-slot contrasts of $-10.2$ and $+17.2$ points. Pooling the
two target types reduces the observed gaps to 10.8 points for Gemini 3.8
($p_H<0.001$) and 7.3 points for Qwen3.5 ($p_H=0.041$). These results show
that opposing target-specific effects can partially cancel in aggregate
profiles.

\begin{figure}[!t]
\centering
\includegraphics[width=\columnwidth]{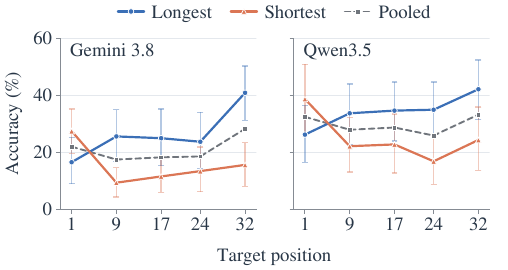}
\caption{Duration accuracy over five orders and two calls, with pointwise 95\% bootstrap CIs.}
\label{fig:crossed_targets}
\end{figure}

\section{RQ3: How Do Effects Change with Input Scale?}\label{sec:scale}
We examine how positional sensitivity changes as the number of candidate
recordings increases. For duration and dysarthria, we construct nested sets
of 5, 8, 12, 20, and 32 candidates from 128 golds per task, balanced across
the two target types. At each scale, we evaluate five relative-position
anchors spanning the first to the last slot, with independent distractor
shuffles.

\subsection{Absolute accuracy changes}

Increasing the candidate count consistently reduces accuracy.
Table~\ref{tab:scale_drop} shows that all 20 model--task--position anchors
decline from 5 to 32 candidates, with losses ranging from 8.6 to 37.5
percentage points. The intermediate scales are not necessarily monotonic,
but the overall degradation is consistent across models, tasks, and
sampled positions.

The degradation, however, is not always uniform across positions.
On dysarthria, Qwen3.5 degrades much less at the first position than at
the remaining anchors, whereas Gemini 3.8 shows a more uniform decline.
Thus, increasing input scale can either preserve or reshape the positional
profile, depending on how unevenly performance degrades across positions.

\begin{table}[!t]
\centering
\caption{Accuracy changes from 5 to 32 candidates at five nominal relative-position anchors (percentage points, $32-5$). All entries are point estimates.}
\label{tab:scale_drop}
\small\setlength{\tabcolsep}{4pt}
\renewcommand{\arraystretch}{1.1}
\begin{tabular}{@{}lrrrrr@{}}
\toprule
Model & $0$ & $1/4$ & $1/2$ & $3/4$ & $1$ \\
\midrule
\rowcolor{black!10}[0pt][0pt]
\multicolumn{6}{@{}l@{}}{\textit{Duration}} \\
Gemini 3.8 Flash & $-21.9$ & $-21.9$ & $-24.2$ & $-26.6$ & $-23.4$ \\
Qwen3.5 Omni Plus & $-37.5$ & $-23.4$ & $-34.4$ & $-29.7$ & $-33.6$ \\
\addlinespace[3pt]
\rowcolor{black!10}[0pt][0pt]
\multicolumn{6}{@{}l@{}}{\textit{Dysarthria}} \\
Gemini 3.8 Flash & $-22.7$ & $-25.0$ & $-18.8$ & $-18.0$ & $-25.8$ \\
Qwen3.5 Omni Plus & $-8.6$ & $-26.6$ & $-29.7$ & $-23.4$ & $-31.2$ \\
\bottomrule
\end{tabular}

\end{table}

\subsection{Relative positional advantages}

\begin{figure}[!t]
\centering
\includegraphics[width=\columnwidth]{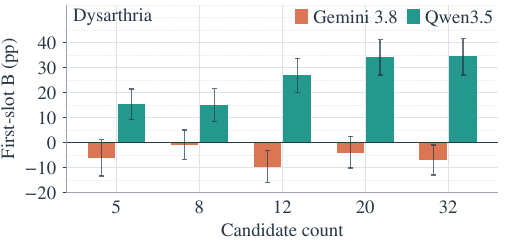}
\caption{Dysarthria first-slot contrast $B$ across candidate counts. Whiskers show pointwise 95\% paired bootstrap intervals.}
\label{fig:scale}
\end{figure}

This uneven degradation can also change relative positional advantages.
Fig.~\ref{fig:scale} shows that Qwen3.5 develops a substantially stronger
first-slot advantage on dysarthria as the candidate count increases,
whereas Gemini 3.8 remains comparatively stable. Importantly, this larger
relative advantage reflects slower degradation at the first position,
rather than improved absolute accuracy.

Consistent with this pattern, a significant scale--position interaction
is found only for Qwen3.5 on dysarthria ($p_H<0.001$). The paired
model-difference test is also significant for dysarthria, but not for
duration. These results suggest that increasing input scale does not
uniformly amplify positional sensitivity: in some settings it primarily
reduces accuracy, while in others it also reshapes the positional profile.

\section{Conclusion}
In this paper, we characterize positional sensitivity in multi-audio reasoning across
models, tasks, target criteria, and input scales. Target position
affects performance, but the direction and magnitude of
these effects vary across tasks and targets rather than following a single
shared pattern. As the number of candidates increases, absolute accuracy
consistently declines, while changes in relative positional effects remain
model- and task-dependent. Positional sensitivity
is a conditional property of multi-audio reasoning rather than a
global bias. Multi-audio evaluation should therefore account for input
position together with the comparison setting and input scale.

\clearpage

\section{Acknowledgments}
This work was supported by Alibaba Research Intern Program. We would like to thank the Qwen Team at Alibaba Token Hub (ATH), Alibaba Group, for providing the computational resources and foundation models (Qwen) used in this research.
\nocite{lostmiddle}
\bibliographystyle{IEEEbib}
\bibliography{references}
\end{document}